\documentclass[useAMS,usenatbib,aas_macros,fleqn,hidelinks,letterpaper]{mn2e}

\usepackage{epsfig}
\usepackage{color}
\usepackage{amssymb}
\usepackage{ams math}
\usepackage[greek,english]{babel}
\usepackage{hyperref}
\usepackage{aas_macros}
\usepackage{multirow,tabularx}

\title{The colour distribution of galaxies at redshift five}
\author[A. B. Rogers et al.]{\parbox\textwidth{A. B. Rogers $^{1}$\thanks{E-mail: abr@roe.ac.uk}, 
R. J. McLure$^{1}$, 
J. S. Dunlop$^{1}$,
R. A. A. Bowler$^{1}$, 
E. F. Curtis-Lake$^{1}$, 
P. Dayal$^{1}$,
S. M. Faber$^{2}$,
H. C. Ferguson$^{3}$,
S. L. Finkelstein$^{4}$,
N. A. Grogin$^{3}$,
N. P. Hathi$^{5}$,
D. Kocevski$^{6}$,
A. M. Koekemoer$^{3}$,
P. Kurczynski$^{7}$
} \\\\
$^{1}$SUPA\thanks{Scottish Universities Physics Alliance}, Institute for Astronomy, University of Edinburgh, Royal Observatory, Edinburgh EH9 3HJ, UK \\
$^{2}$University of California Observatories/Lick Observatory, University of California, Santa Cruz, CA 95064, USA \\
$^{3}$Space Telescope Science Institute, 3700 San Martin Dr., Baltimore, MD 21218, USA \\
$^{4}$The University of Texas at Austin, Austin, TX 78759, USA \\
$^{5}$Aix Marseille Universit\'{e}, CNRS, LAM (Laboratoire d'Astrophysique de Marseille) UMR 7326, 13388, Marseille, France \\
$^{6}$Department of Physics and Astronomy, University of Kentucky, Lexington, KY 40506, USA \\
$^{7}$Department of Physics and Astronomy, Rutgers University, Piscataway, NJ 08854, USA
}
\begin{document}

\date{Accepted 2014 March 18. Received 2014 February 21; in original form 2013 December 17}
\pagerange{\pageref{firstpage}--\pageref{lastpage}} \pubyear{2014}

\maketitle

\label{firstpage}

\begin{abstract}
We present the results of a study investigating the rest-frame ultra-violet (UV)
spectral slopes of redshift $z\approx 5$ Lyman-break galaxies (LBGs). 
By combining deep {\it Hubble Space Telescope} imaging of the
CANDELS and HUDF fields with ground-based imaging from the UKIDSS Ultra Deep Survey (UDS), we have produced a
large sample of $z\approx 5$ LBGs spanning an unprecedented factor of
$>100$ in UV luminosity. Based on this sample we find a clear colour-magnitude relation (CMR) at
$z\approx 5$, such that the rest-frame UV slopes ($\beta$) of brighter
galaxies are notably redder than their fainter counterparts. 
We determine that the $z\approx 5$ CMR is well described by a linear
relationship of the form: ${\rm d}\beta = (-0.12\pm0.02){\rm d}M_{\rm UV}$,
with no clear evidence for a change in CMR slope at faint magnitudes (i.e. $M_{\rm UV}\geq -18.9$).
Using the results of detailed simulations we are able, for the first
time, to infer the intrinsic (i.e. free from noise) variation of galaxy colours around the
CMR at $z\approx 5$. 
We find significant ($12\sigma$) evidence for intrinsic colour variation in the sample as a whole.
Our results also demonstrate that the width of the intrinsic UV slope distribution of $z\approx 5$ galaxies increases from $\Delta \beta \simeq0.1$ at $M_{\rm UV}=-18$ to $\Delta \beta \simeq0.4$ at $M_{\rm UV}=-21$.
We suggest that the increasing width of the intrinsic galaxy colour distribution and the CMR itself are both plausibly explained by a luminosity independent lower limit of $\beta \approx -2.1$, combined with an increase in the fraction of red galaxies in brighter UV-luminosity bins.
\end{abstract}



\section{Introduction}

The rest-frame ultra-violet (UV) properties of star-forming galaxies at $z\gtrsim3$ can potentially provide a powerful insight into the formation and evolution of galaxies at high redshift.
In common with local star-burst galaxies \citep{Steidel1999,Meurer1999}, the UV light of high-redshift galaxies is
dominated by short-lived massive stars which should provide a good probe of the current metallicity and dust conditions within rapidly evolving stellar populations.

These galaxies are identified by spectral discontinuities at rest-frame $912\,$\AA\ and $1216\,$\AA, so are typically termed Lyman-break galaxies (LBGs, see \citealt{Giavalisco2003}) regardless of how they are selected.
Redward of the Lyman break, the rest-frame UV continuum of star-forming galaxies 
is traditionally parametrized by the spectral index $\beta$, under
the assumption \citep[see][]{Leitherer1999} that the UV continuum can be approximated by a
power-law: $f_\lambda\propto\lambda^\beta$.
Studies of star-forming galaxies at low redshift have revealed a strong
relationship between UV slope and reddening \citep{Meurer1999}, such 
that the spectral slope $\beta$ and differential extinction $E(B-V)$ are often quoted interchangeably.
More recent studies exploiting {\it Herschel} data have shown that the
relationship between $\beta$ and dust attenuation appears to be
already in place by $z\approx 2$ \citep{Reddy2012}, 
even though the youngest galaxies ($<$100 Myr) may follow a different extinction curve to their older counterparts \citep{Reddy2010}.

At higher redshifts ($z\geq3$) the relationship between UV slope, dust
attenuation, stellar population age and metallicity is still unclear.
However, many previous studies have shown that $\beta$ reddens with
decreasing redshift and increasing UV luminosity
\citep[e.g.][]{Meurer1999, Shapley2003, Labbe2007, Overzier2008,
  Wilkins2011, Bouwens2013, Dayal2012}. This trend is often assumed to reflect
increasing dust attenuation at a fixed redshift, and increasing
stellar population age at a fixed luminosity
\citep[e.g.][]{Labbe2007}.

The obvious prediction from the apparent trend between UV slope,
redshift and luminosity is that the bluest galaxies will naturally be found
amongst the faintest detectable galaxies at $z\geq7$. Indeed, initial
analysis of ultra-faint LBGs detected in near-IR {\it HST} imaging of
the Hubble Ultra-deep field (HUDF) suggested a
population of galaxies with very blue UV slopes (i.e. $\beta \leq -3$),
which would require dust free, ultra-low metallicity stellar
populations \citep[][but see \citealt{Finkelstein2010}]{Bouwens2010}. 
However, careful consideration of the
observational biases (e.g. \citealt{Dunlop2012}; \citealt{Finkelstein2012}; Rogers, McLure \& Dunlop 2013; \citealt{Wilkins2011}) and the availability of yet deeper {\it HST} imaging
\citep{Ellis2013, Koekemoer2013} has led to a consensus that the UV
slopes of currently observable $z\approx7$ galaxies are not
significantly bluer than those of moderately young, but
otherwise unremarkable, stellar populations \citep{Dunlop2013, Bouwens2013}.

As a consequence, attention has now turned to understanding the
details of how the relationship between UV slope and luminosity,
hereafter the colour-magnitude relation (CMR), evolves at high
redshift ($z\geq 4$). At present, there is no consensus on the
strength or functional form of the CMR at high redshift,  with the two
largest studies at $z\geq4$, those of \citet[hereafter
F12]{Finkelstein2012} and \citet[hereafter B13]{Bouwens2013}, producing seemingly discrepant results.
Although both studies are based on samples with a reasonable dynamic
range in UV luminosity, selected from high-quality {\it HST} imaging,
B13 claim the discovery of a significant CMR in redshift bins at $z=4, 5, 6,$ \& $7$, whereas
F12 see no significant correlation between $\beta$ and $M_{\rm UV}$
in the same redshift range. In contrast, F12 find that $\beta$ is more
strongly coupled to stellar mass $M_\star$, with more massive galaxies displaying redder UV slopes.

In addition to fundamental questions related to the existence and form
of the CMR at high redshift, constraints on the {\it intrinsic}\footnote{Throughout this work, we refer to the noise-free distribution of colours as `intrinsic', i.e. free of observational effects.} colour distribution of $z\geq 4$ galaxies are clearly of interest.

For instance, at $z<3$, \citet{Labbe2007} found evidence for moderate
intrinsic colour-variation in the blue sequence, which they attributed to
stochastic star-formation histories. However, previous studies
addressing this issue at $z\approx4-5$ have been hampered by a lack of
dynamic range \citep{Castellano2012, Bouwens2012,Wilkins2011} and no luminosity-dependent trend is currently clear.

The initial aim of this paper is to combine the strength of deep,
small area, {\it HST} imaging with shallower, but wide area,
ground-based imaging from the UKIDSS Ultra Deep Survey (UDS) to
provide a large sample of
$z\approx 5$ LBGs spanning an unprecedented dynamic range in UV
luminosity ($\approx 5$ magnitudes). We focus entirely on $z\approx 5$
galaxies because this is the highest redshift for which it is possible
to consistently select large samples of galaxies free from Ly~$\alpha$
contamination using the deep
$z$-band imaging available across our
{\it HST} and ground-based datasets. Based on the techniques developed
in \cite{Rogers2013}, we use bias-free measurements of $\beta$ to
provide the best available constraints on the form of the $z\approx 5$ CMR, before proceeding to exploit the results of detailed
simulations to investigate the {\it intrinsic} galaxy colour distribution
as a function of luminosity.

The remainder of this paper is organised as follows.
In Section 2 we describe the selection of our $z\approx5$ galaxy sample
from our chosen {\it HST} and ground-based datasets.
In Section 3 we briefly recount our method for measuring rest-frame UV
colours and present our constraints on the $z\approx 5$ CMR. In Section
4 we describe the detailed simulations which were necessary to
accurately quantify the contribution of photometric uncertainties to the observed galaxy colour distribution.
Based on the results of these simulations, in Section 5 we present our
measurement of the intrinsic scatter in the galaxy colour distribution
as a function of UV luminosity. Our conclusions are summarised in
Section 6. Throughout the paper we quote magnitudes in the AB system
\citep{Oke1983} and assume a cosmology 
with $\Omega_0=0.3$, $\Omega_\Lambda=0.7$, $H_0=70$~km s$^{-1}$~Mpc$^{-1}$.
We refer to \emph{HST}'s ACS F435W, F606W, F775W, F814W, F850LP and WFC3/IR
F098M, F105W, F125W, F140W and F160W filters as $B_{435}, V_{606},
i_{775}, I_{814}, z_{850}, Y_{098}, Y_{105}, J_{125}, JH_{140}$ and $H_{160}$
respectively.


\section{Data and sample}
\begin{figure*}
\includegraphics[width=7.0in]{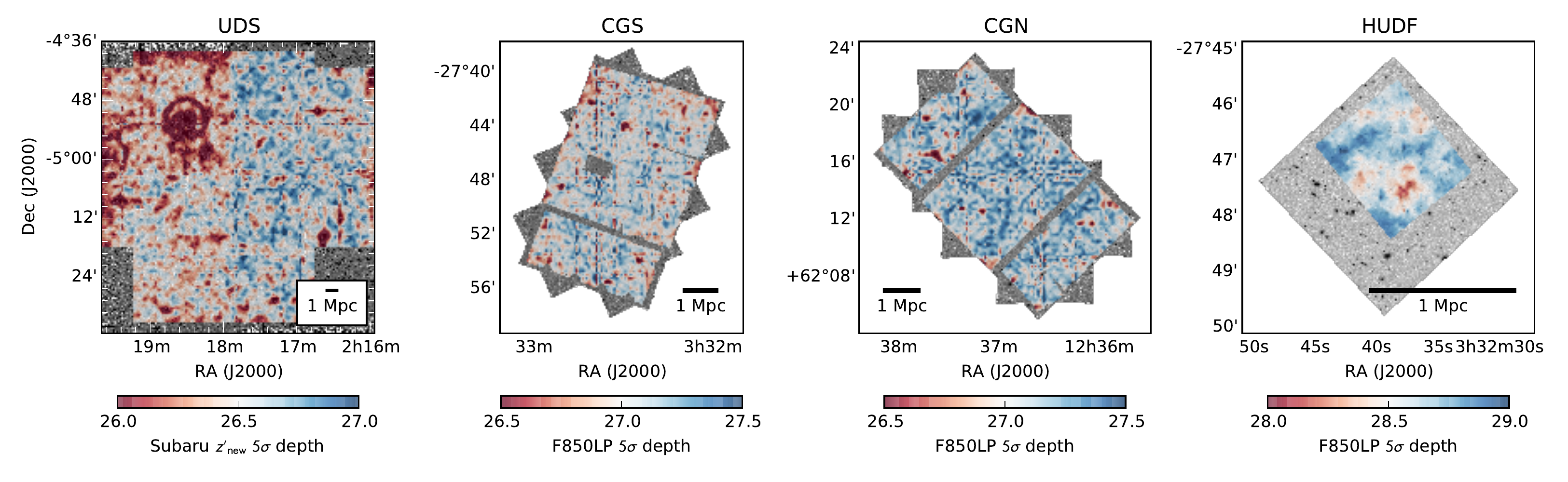}
\caption{The four fields analysed in this study. 
In each, the greyscale image shows the $z$-band imaging in which the galaxies are detected, while the colour-map shows the local $5\sigma$ $z$-band depth from which SNR cuts are applied.
Depths are computed at nodes of a $100\times100$ grid, based on each node's 200 nearest source-free apertures (see Section \ref{sect:depths} for a discussion).
The footprint of each depth-map defines our search area, i.e. the area in which imaging exists in all wavebands.
The scalebar denotes a \emph{physical} size of approximately 1~Mpc at $z=5$.}
\label{fig:maps}
\end{figure*}

In this section we describe the selection of our $z\approx5$ galaxy sample.
The sample is selected from four survey fields: the UKIRT Infrared Deep Sky Survey (UKIDSS) Ultra Deep Survey (UDS), the Cosmic Assembly Near-infrared Deep Extragalactic Legacy Survey (CANDELS) programme's coverage of the Great Observatories Origins Deep Survey (GOODS) North and South fields, and the Hubble Ultra Deep Field (HUDF).
Summary details of the four fields and their respective $z\approx5$ LBG samples are given in Tables \ref{tab:fields} and \ref{tab:sample}.

\subsection{Description of imaging}
\begin{figure}
\centering
\includegraphics[width=3.1in]{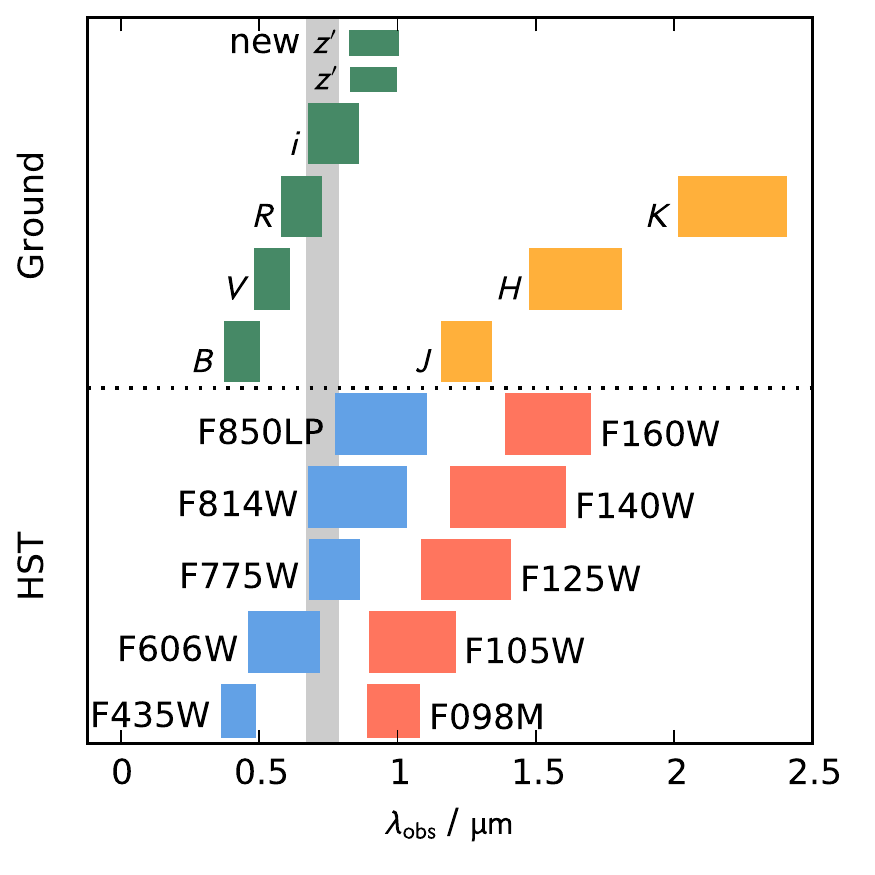}
\caption{The photometric filter bandpasses for the images used in this work are shown by the coloured regions.
Blue and red regions show the \emph{HST}'s ACS and WFC3/IR filters, while green and yellow regions show the Subaru and UKIRT filters used in the UDS.
The vertical, grey region denotes the wavelength range in which the 1216~\AA\ Lyman break is observed at $4.5<z<5.5$.}
\label{fig:filters}
\end{figure}
Here, we briefly describe the available imaging in each field.
The fields themselves are shown in Fig. \ref{fig:maps}, while the band-passes of the filters are shown in Fig. \ref{fig:filters}.
\subsubsection{UKIDSS Ultra Deep Survey (UDS)}
\label{sub:uds}
The UDS is covered by public Subaru $B,V,R,i,z'$-band \citep{Furusawa2008} and UKIRT $J,H,K$-band imaging\footnote{The images we use are from DR10. See \url{http://www.ukidss.org}}, with a co-imaged area of 0.6 square degrees.
In addition, we have made use of additional $z'$-band imaging taken after Subaru's SuprimeCam was refurbished with CCDs with improved red sensitivity.
For the remainder of this work, the public $z'$ band is referred to as $z_{\rm old}$ and the new, deeper, proprietary $z'$ band as $z_{\rm new}$.

\subsubsection{CANDELS GOODS-N (CGN) and CANDELS GOODS-S (CGS)}
The two CANDELS GOODS fields together provide 283~arcmin$^2$ of \emph{HST} ACS and WFC3/IR imaging.
The survey and data reduction are described by \citet{Grogin2011} and \citet{Koekemoer2011} respectively.
In each field the WFC3/IR imaging consists of a deep central region, flanked by two `wide' strips.
In CGS, one of these wide strips is provided by the Early Release Science field (ERS, \citealp{Windhorst2011}), which features deep $Y$-band coverage through the $Y_{098}$ filter rather than the $Y_{105}$ filter which is employed over the rest of the CGN and CGS fields (see Table \ref{tab:fields}).

\subsubsection{Hubble Ultra Deep Field (HUDF)}
The HUDF has been imaged by multiple programmes, most recently in the near-IR by the UDF12 campaign \citep{Ellis2013,Koekemoer2013}.
We have analysed the UDF12 near-IR imaging in tandem with the ACS optical $B_{435},V_{606},i_{775},z_{850}$-band imaging provided by \citet{Beckwith2006} and the more recently assembled $I_{814}$ imaging provided as part of the XDF \citep{Illingworth2013}.
For this work, we searched only the area covered by the deepest WFC3/IR imaging ($\approx4$~arcmin$^2$).

\subsection{Photometry}

\begin{table*}
\begin{tabular}{l c c c c c c c c c c c c}
\hline
Field    & Area/arcmin$^2$  & SNR cut &   \multicolumn{10}{c}{$5\sigma$ depth/AB mag} \\
 &    & & $B_{435}$ & $V_{606}$ & $i_{775}$ & $I_{814}$ & $z_{850}$ & ${Y_{098}}^a$ & ${Y_{105}}$ & $J_{125}$ & $JH_{140}$ & $H_{160}$\\
\hline
HUDF & 4  & 5 & 29.2 & 29.6 & 29.2 & 28.4 & 28.6 & -- & 29.5 & 29.2 & 29.2 & 29.3  \\
CGS & 143  & 8 & 27.6 & 27.8 & 27.3 & -- & 27.0 & 27.1 & 27.5 & 27.5 & -- & 27.3 \\ 
CGN & 140  & 8 & 27.6 & 27.8 & 28.4 & -- & 27.2 & -- & 26.9 & 27.1 & -- & 27.0\\ 
\\
\cline{4-13}
&&& $B$ & $V$ & $R$ & $i$ & $z_{\rm old}$ & $z_{\rm new}$ & $J$ & $H$ & $K_s$\\
\cline{4-13}
UDS & 2701 & 12 & 27.8 & 27.5 & 27.2 & 27.1 & 26.1 & 26.5 & 25.6 & 25.1 & 25.3\\
\\\hline
\textsc{total:} & 2988 \\
\hline
\end{tabular}
\caption{Summary of the fields used in this work.
Columns 1 and 2 list the field names and the corresponding area of each used to search for $z\approx5$ galaxies.
Column 3 lists the $z$-band SNR threshold adopted for detecting galaxies in each field.
Columns 4--13 list the depth of imaging in each field. 
Quoted depths are medians across the field, measured in circular apertures of diameter 0.6 arcsec (\emph{HST}) or 1.8 arcsec (UDS); however, for selection, local depth measurements are used (see Section \ref{sect:depths}). 
The CANDELS CGS and CGN have variable depth NIR imaging, typically ranging from 27 -- 28 mag, while the UDS imaging varies in the $z_{\rm new}$ selection band as shown in Fig. \ref{fig:maps}.
$^a$ -- $Y_{098}$ imaging is used in place of $Y_{105}$ for the northern strip (ERS; 44 arcmin$^2$) of the GOODS-S data.}
\label{tab:fields}
\end{table*}

Fixed-diameter circular apertures were used to construct photometric catalogues from each image.
In the UDS imaging, 1.8-arcsec diameter apertures were used in all bands, enclosing $\approx80$\% of a point source's flux.
While a point source is a reasonable approximation to a $z\approx5$ galaxy in the ground-based data, \emph{HST} allows many of the brighter galaxies to be resolved. 
This is advantageous, as it allows stars to be easily distinguished from galaxies using their measured half-light radii ($r_{1/2}$).
However in small PSF-matched apertures, broader sources lose a larger fraction of their light in short wavebands, resulting in a red colour bias for extended objects.
There are various options to alleviate this: 
PSF homogenisation, which relies on a well-constrained transfer function to match the PSF of each image to that of the $H_{160}$ data (which has the poorest spatial resolution);
a measured-size dependent correction to the aperture photometry, which relies on well-measured half-light radii; 
or the adoption of sufficiently large apertures that realistic high-redshift galaxy sizes ($r_{1/2}$ $\lesssim$ 1.5~kpc) have a negligible impact, at the expense of image depth.
For this work, we assume the third approach and use apertures of diameter 0.6 arcsec for all \emph{HST} photometry.
These apertures enclose a sufficient fraction of the total light in all bands that biases in the UV slope are at the level of $\left|\delta\beta\right|\lesssim0.2$ (see Section \ref{sec:sims}).

\subsection{Image depths}
\label{sect:depths}

Given the variable exposure-time maps of each survey field, and the importance of consistent signal-to-noise cuts across our sample, the SNR of the photometry for each candidate, in each filter, has been computed relative to the local image depths.
To do so we created an object (segmentation) mask for each mosaic using \textsc{sextractor} v2.8.6 \citep{Bertin1996}, set to mask out any area where two or more pixels rise above 1.4$\sigma$, and placed non-overlapping apertures across the remaining source-free sky regions.
The local depth at a given point on the mosaic was then measured by analysing the nearest 200 source-free apertures and computing the $5\sigma$ depth using the Median Absolute Deviation (MAD) statistic,
which yields the standard deviation of the distribution of fluxes by $\sigma \approx 1.4826 \times \textrm{MAD}$.
The MAD is ideal for these measurements as it is robust against very non-Gaussian distributions.
The depth maps shown in Fig. \ref{fig:maps} were created by computing local depths at nodes of a grid across the images, while the values quoted in Table \ref{tab:fields} are medians over each field.

\subsection{Selection of \lowercase{$\bmath{z\approx5}$} galaxies}

Within each field, candidates were initially detected using \textsc{sextractor} in dual-image mode, detecting in the $z$-band image and measuring from all others in fixed, circular apertures as discussed above.
SNR thresholds were then applied in the $z$ band, to remove sources detected at low significance, and in the $B$ band, since both the Lyman break ($\lambda_{\rm rest}=$1216~\AA) and limit (912~\AA) of a genuine $z\approx5$ galaxy would lie redward of the $B$ band.
To do so, the catalogues were first reduced in size by keeping only candidates with $z$-band detections brighter than the $3\sigma_{\rm global}$ depth (in the HUDF) or the $5\sigma_{\rm global}$ depth in CGN, CGS and the UDS, where $\sigma_{\rm global}$ is the median of local depths across a given image.
At the same time, the catalogues were pruned of any object with a $B$-band detection at the $2\sigma_{\rm global}$ level.
For the remaining candidates, local depth estimates were computed in each band, using the procedure outlined in Section \ref{sect:depths}.
A second cull of $B\geq2\sigma_{\rm local}$ detected sources removed the few contaminants lying in deeper parts of the image.
Finally, refined $z$-band signal-to-noise cuts were imposed at local $5\sigma,8\sigma,8\sigma,12\sigma$ thresholds for candidates in the HUDF, CGN, CGS and UDS (as per Table \ref{tab:fields}).
These thresholds were chosen as compromises between sample size and data quality, which is itself dependent on the homogeneity of depths and the number of bands probing the rest-frame UV.
The primary driver of these chosen thresholds was to ensure that, regardless of the field, the bands to be used for measuring $\beta$ would all have high signal-to-noise for sources with a reasonable (i.e. flat) spectral slope.
In the UDS for example, the $J$ and $H$ bands, which together with the $z$ band probe the rest-frame UV, have depths 0.9 and 1.4 mag shallower than the $z_{\rm new}$ selection band.
A flat-spectrum ($\beta=-2$) source, detected at $5\sigma$ in the $z_{\rm new}$ band, would have $J$- and $H$-band photometry with SNR of only $\approx2$.
Raising the $z$-band selection threshold counters this problem.

\subsubsection{Photometric redshift analysis}
\label{sec:photzs}

\begin{figure}
\centering
\includegraphics[width=3.5in]{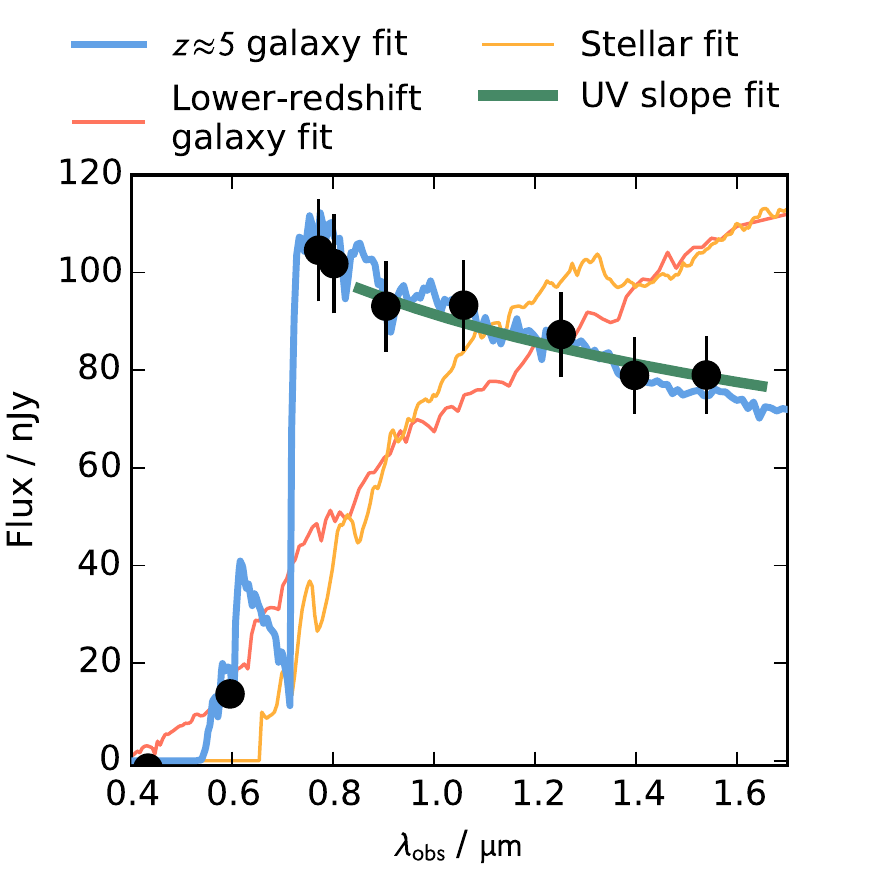}
\caption{The spectral energy distribution of an example $z\approx5$ galaxy in the HUDF is shown by black points, in the observed frame.
The lines show the various fits used to analyse the galaxy.
The error bars are inflated to at least 10\% of the flux for the fitting (as shown here; and see Section \ref{sec:photzs}).
In blue is the best-fitting primary redshift solution.
The possibility of it being a contaminant is ruled out by the poor lower-redshift galaxy and stellar fits (red and yellow).
The rest-frame UV spectral index $\beta$ is deduced by fitting with a power-law model (green) to the observed near-IR data.}
\label{fig:fitting}
\end{figure} 

Redshift $4.5\leq z\leq5.5$ galaxies were selected using the photometric redshift code \textsc{lephare} \citep{Arnouts1999, Ilbert2006}, adopting the `COSMOS' galaxy models of \citet{Ilbert2006} and galactic dwarf star templates from the SpeX library\footnote{\url{http://pono.ucsd.edu/~adam/browndwarfs/spexprism}. For further details see Acknowledgements.}.
The `COSMOS' spectral energy distributions (SEDs) include elliptical, spiral and star-burst templates, which were fit over the redshift range (0, 10) allowing dust reddening of $E(B-V)\leq1.5$ with a \citet{Calzetti2000} reddening law.
While our local depth estimates are robust, we ensure the entire observed SED contributes to the fit by imposing \emph{minimum} flux errors of 10\%\footnote{In practice, these minimum errors were adopted in the NIR bands for around 10\% of candidates brighter than $z>5\sigma$ in the UDS, to 50\% of candidates in the HUDF.}.

Candidate $z\approx5$ galaxies were thus selected, regardless of how marginally the $z\approx5$ photometric redshift solution was preferred to any secondary solution.
As discussed at length by \citet{Rogers2013} and \citet{Dunlop2013}, introducing any form of $\chi^2(\rm {primary}) - \chi^2(\rm secondary)$ threshold inevitably biases the selection against faint, intrinsically (or photometrically scattered) red galaxies, although this effect is mostly alleviated for galaxies detected at SNR~$\gtrsim8\sigma$ \citep{Dunlop2012}.
In reality, given our chosen redshift and signal-to-noise windows and large apertures, lower redshift models typically struggle to fit the shape of the observed purported Lyman breaks.
Fig. \ref{fig:fitting} shows an example of the model fitting procedure.
Genuine high-redshift galaxies were considered to be those for which the primary galaxy model SED (with an assumed four degrees-of-freedom) was acceptable at the $2\sigma$ level, i.e. $\chi^2\leq 11.3, 9.7, 7.9$ in the HUDF, UDS and CGS/N fields respectively.
Potential stellar contaminants were rejected if the best stellar fit was acceptable at $2\sigma$ and if the source's $z$-band half-light radius $r_{1/2}$ was consistent with being stellar at that luminosity.

Finally, the imaging and SED fits for each candidate were visually inspected.
Having survived the selection procedure thus far, lone objects were only deleted if they were deemed to lie too close to the image edge, or if their SED was acceptably stellar but their proximity to another source led them to be measured with an excessive $r_{1/2}$.
Candidate galaxies were also removed from the sample if they were close enough to a large foreground object that the photometry or background subtraction were likely to be insecure, or if they were blended with another smaller object with a significantly different photometric redshift. 
In deriving those photometric redshifts, \textsc{sextractor} was used to determine the photometry for each source with the other(s) masked.
Additionally, blending in the low-resolution (ground-based) UDS imaging was checked by cross-matching our catalogue to that of \citet{Galametz2013,Galametz2013cat}, which was derived from the \emph{HST} CANDELS UDS imaging \citep{Grogin2011,Koekemoer2011}, which covers a small portion of the ground-based UDS (202~arcmin$^2$ of the total 2701~arcmin$^2$).
Of our 27 UDS candidates also in the CANDELS catalogue, only one is revealed as having two distinct components in the \emph{HST} imaging.
Based on a CANDELS UDS photometric redshift catalogue \citep{Galametz2013,Fontana2006}, the two components have $z=4.61$ and $z=4.65$, in good agreement with $z=4.75$ from the ground-based imaging alone, and are visually connected in the \emph{HST} images suggesting the two components are in fact a single source.
Assuming that the CANDELS UDS region is representative of the whole UDS field, the fraction of ground-based UDS galaxy candidates which in the other fields would have been treated as multiple individual sources is expected to be $\lesssim5$\%.

Having applied the above selection procedure, our sample consists of 584 LBG candidates at $z\approx5$, selected over 2988 arcmin$^2$.
The sample covers a long luminosity baseline of $-22.5<M_{1500}<-17.5$, large enough to provide strong leverage on the CMR.
A summary of the sample's properties, broken down by field, is given in Table \ref{tab:sample}.

\begin{table}
\begin{tabular}{l c c c c c c}
\hline
Field & $N$ &$\langle z\rangle$ & $\langle M_{\rm UV}\rangle$ & $\langle\beta\rangle$ & \multicolumn{2}{c}{Mean$(\delta\beta)$} \\
& & & & & (data) & (sim) \\
\hline
HUDF &  33 & 5.1 & $-$18.5 & $-$2.04$\pm$0.05 & 0.26 & 0.26\\
CGS & 112 & 4.9 & $-$20.2 & $-$1.82$\pm$0.04 & 0.27 & 0.30\\
CGN & 163 & 5.0 & $-$20.1 & $-$1.90$\pm$0.04 & 0.29 & 0.29\\
UDS & 276 & 4.9 & $-$21.2 & $-$1.90$\pm$0.03 & 0.39 & 0.36\\
\hline 
\end{tabular}
\caption{Sample properties of galaxies from the four field analysed in this work (column 1). 
Column 2 lists the number of galaxies selected from each field.
Columns 3--5 list the mean redshift, mean absolute magnitude, and the mean UV slope and its standard error. 
The final two columns list the mean error on the $\beta$ measurement for an individual galaxy from the data and, for comparison, the simulations described in Section \ref{sec:sims}.}
\label{tab:sample}
\end{table}

\subsection{Selection method validation}
To check our sample selection method, we have compared our CANDELS GOODS-S (CGS) sample to other catalogues.
Comparing to the ESO GOODS/CDF-S Spectroscopy master catalogue\footnote{\url{http://www.eso.org/sci/activities/garching/projects/goods/MasterSpectroscopy.html}}, we find 23 sources with spectroscopic redshifts from \citet{Vanzella2008}, and three with spectroscopic redshifts from \citet{Popesso2009} and \citet{Balestra2010}.
All 26 spectroscopic redshifts lie in the range $4.4<z_{\rm spec}<5.6$, in excellent agreement with our $4.5<z_{\rm phot}<5.5$ defined selection window.
The photometric redshift accuracy is $\sigma[(z_{\rm spec} - z_{\rm phot})/(1+z_{\rm spec})]=0.032$.
Of our 112 CGS galaxies, 110 have also been studied by Dahlen et al. (2014, in prep.), who compared photometric redshift estimates from 11 different codes \citep{Dahlen2013}.
Comparing to their results, we estimate that our lower-redshift interloper contamination fraction is around 5\%.
This is due to our inclusion of candidates with good high-redshift solutions, but which are only marginally distinguished from lower-redshift solutions.
This estimate is also in line with a comparison of the 27 of our UDS galaxies which fall within the CANDELS UDS \emph{HST} coverage: using the CANDELS photometry and independent photometric redshifts \citep{Galametz2013,Galametz2013cat,Fontana2006} reveals only one of the 27 galaxies to have a preferred lower-redshift fit.

Fortunately, as we shall see in Section \ref{sec:interlopers}, the exclusion of these potential contaminants, which span a wide range of luminosities, does not affect the results that follow.

\subsection{Measuring $\bmath {M_{1500}}$}
Absolute UV magnitudes ($M_{\rm UV}=M_{1500}$) were determined using a top-hat filter centred on 1500~\AA\ in the rest-frame of the best-fitting SED template.
To make our results comparable to other studies, the absolute magnitudes were boosted by around $0.2$~mag, to account for the still incomplete enclosure of a PSF's flux within our already large apertures.
The luminosity distribution of our complete sample is shown in the top panel of Fig. \ref{fig:cmr}.
By design, this definition of $M_{1500}$ does not include any correction for dust attenuation; it is the `observed' absolute magnitude.
This choice serves to avoid imparting an artificial colour--magnitude relation onto the data.
For instance, the simplest correction would be to follow the \citet{Meurer1999} relation, which yields a dust attenuation correction $A_{1600}$ based on the observed colour $\beta$:
a low $\beta$ suggests a small $A_{1600}$, so only a minimal correction is required on $M_{1500}$.
While less direct, the same must be true if the dust attenuation is estimated by SED fitting, since the SED fit has essentially no other information from which to constrain the dust attenuation.
Thus, had we adopted such a correction, red galaxies would be luminosity-corrected by a larger degree than blue galaxies at the same \emph{observed} luminosity.
Under the null hypothesis that $M_{1500}$ and $\beta$ are uncorrelated, or are only weakly correlated, the result is to skew the observed $\beta$ vs. $M_{1500,\,\rm{obs.}}$ space, generating a fake dependency of $\beta$ on $M_{1500,\,\rm{corrected}}$.


\section{Measurement of UV slopes and the colour--magnitude relation}
In this section, we use our sample to constrain the colour--magnitude relation at $z\approx5$.

\subsection{Measuring the UV slope}
Following \citet{Rogers2013} and \citet{Dunlop2013}, we fitted for $\beta$ at fixed redshift using a variety of power-law SED models.
Adopting the photometric redshift $z_{\rm phot}$ derived at the selection stage, we re-ran \textsc{lephare} treating $z_{\rm phot}$ as a spectroscopic redshift.
The models were pure power-law models ($f_{\lambda} \propto \lambda^\beta$), truncated shortward of 912~\AA, with no dust attenuation allowed.
Attenuation due to the IGM is provided by \textsc{lephare}.
An example of these power-law fits is shown in Fig. \ref{fig:fitting}.
Minimum photometric errors of 10\% are still used; removing these has little effect, with $\beta$ changing by $<0.1$.
The fitting procedure yields, for each object, a distribution $\chi^2(\beta_i)$ for a finely gridded set of templates $-8<\beta_i<+5$.
The best-fitting UV slope was taken to be that which minimized $\chi^2$, and the error was obtained by finding the values of $\beta$ for which $\Delta\chi^2=1$ from the $\chi^2$ minimum (where the $\chi^2$ distribution is minimized over the SED normalisation).
This procedure excludes the colour uncertainty induced by the photometric redshift uncertainty, but in the vast majority of cases this is minimal since the Lyman break falls a good way blueward of the $\beta$-measuring wavebands (see Fig. \ref{fig:filters}).
In the UDS, we used the two $z$ bands, $J$ and $H$ to fit $\beta$.
In the CANDELS fields, we used $z_{850}, Y_{098|105}, J_{125}$, and $H_{160}$, and in the HUDF, we used $z_{850}, Y_{105}, J_{125}, JH_{140}$, and $H_{160}$. 

\subsection{Linear fits to the colour--magnitude relation}
\begin{figure}
\includegraphics[width=3.3in]{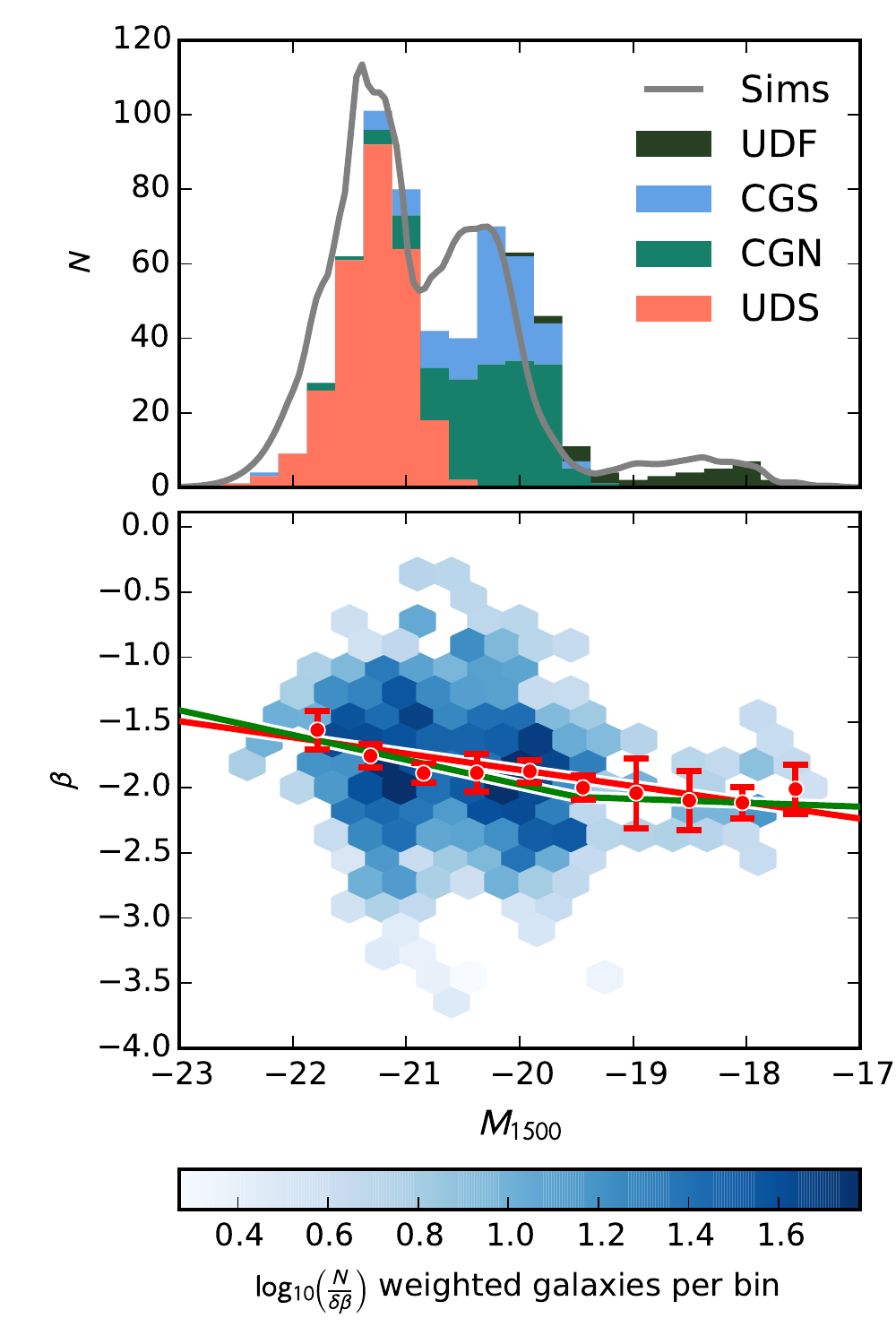}
\caption{\emph{Top:} the luminosity distribution of the sample used to constrain the colour--magnitude relation, shown as a stacked histogram, split by field.
The grey line shows the expected distribution based on the $z=5$ luminosity function and the size and selection function of each field (see later simulations).
\emph{Bottom:} the density map shows the combined sample in colour--magnitude space.
Included as a visual aid only, the red circles are binned means, and their error bars are $2\times$ the standard error on the mean in each bin (inflated for visibility).
These error bars are driven by photometric scatter, so do not directly constrain any intrinsic variation that may exist in the colours. 
The red line (the linear CMR) is a weighted fit to all the individual galaxies, with slope $-0.12$.
Both the density map and fit are weighted by the uncertainty, $\delta\beta$, on each galaxy's $\beta$ colour.
The green line is the two-component piecewise-linear fit reported by \citet{Bouwens2013}.
Their single linear relation is in near-perfect agreement with ours, so is not shown.
}
\label{fig:cmr}
\end{figure}

A simple linear fit to the colour--magnitude distribution of the entire galaxy sample, with each galaxy weighted by its colour error $1/\delta\beta$, yields a slope
\begin{equation*}
\frac{{\rm d}\beta}{{\rm d}M} = -0.12\pm 0.02,
\end{equation*}
and a zeropoint of
\begin{equation*}
\beta(M_{\rm 1500}=-19.5)=-1.93 \pm 0.03.
\end{equation*}
Fig. \ref{fig:cmr} shows this relation fitted to the sample.
These values are in excellent agreement with the results of B13, who used space-based data alone.
This parametrization does differ from the results of F12, who found a much weaker correlation, but our relation is still statistically consistent with their binned data points.
As F12 illustrated, restricting their faintest bin to galaxies from the HUDF alone (which moves the bin fainter) also yields a stronger relation, bringing it further in line with this work.

Following the suggested existence of a piecewise-linear relation by B13, whereby ${\rm d}\beta/{\rm d}M$ is steeper at the bright end (e.g. $M_{1500}<-18.9$) than the faint end, Fig. \ref{fig:cmr} also shows their two-component broken linear fit.
Our data show no clear evidence to support the broken power-law CMR (see the binned mean results in Fig. \ref{fig:cmr}).
Fixing the break at B13's suggestion of $M_{1500}=-18.9$ and fitting the bright- and faint-end slopes of our data does yield values similar to those reported by B13; of $-0.16\pm0.02$ and $-0.04\pm0.12$ respectively.
However, given the uncertainties, there is clearly no significant evidence for the CMR being non-linear.
Moreover, even the qualitative nature of this result is heavily dependent on the exact choice of the break luminosity.
For example, floating the break luminosity fails to yield any meaningful constraint, and fixing the break instead at $M_{1500}=-19.5$ results in a steeper faint- than bright-end CMR gradient.
In summary, we confirm the existence of a significant CMR with our sample.
However at least in this redshift window (B13 studied a range of redshifts) we lack strong evidence to either confirm or refute the existence of a characteristic luminosity at which the colour--magnitude relation changes gradient.

\subsubsection{The impact of potential interlopers}
\label{sec:interlopers}
Of the 584 galaxies in the combined sample, 15 have secondary redshift solutions (mostly at $z_{\rm sec}<1$) which are acceptable at the $2\sigma$ level as defined in Section \ref{sec:photzs}.
Removing these 15 galaxies does not alter the measured CMR parameters beyond the stated confidence intervals:
$\beta({\rm no\;interlopers})=(-0.13\pm0.02)M_{1500}-(1.96\pm0.03)$.

\section{Image simulations}
\label{sec:sims}
In this section we describe the creation and injection of simulated galaxies into the images, and explain how their detection and selection efficiencies compare to the real galaxy sample.
\subsection{Model galaxies}
Simulated galaxies were defined on a grid spanning $\beta, M_{\rm UV}$ and $z$, in order that the selection probability could be computed for any galaxy with a given intrinsic redshift, luminosity and colour, and so that mappings could be made between intrinsic and observed values of $\beta$.
The luminosity distribution from which they were drawn conforms to the simple redshift-evolving luminosity function of \citet[section 2.7]{McLure2013}.
This evolving luminosity function was needed since, to simulate some scattering between redshift bins, we input galaxies over the redshift range $4<z<6$, through which the luminosity function evolves significantly.

Each model's SED and $M_{\rm UV}$ were derived from a BC03 \citep{Bruzual2003} stellar population synthesis model with metallicity $0.2$~Z$_\odot$, a declining star formation history with $\tau=0.03$~Gyr and a \citet{Chabrier2003} IMF. 
The galaxy population was defined with a uniform distribution of galaxies in the $\beta$ dimension, achieved by mapping $\beta$ to pairs of stellar population age $t$ and  \citet{Calzetti2000} dust reddening $E(B-V)$ (see fig. 3 of \citealt{Rogers2013}). 
While the entire range of $\beta$ could have been reproduced by modifying $E(B-V)$ at fixed $t$, the age evolution was imposed to provide a more physically motivated model at each $\beta$. 
Galaxies were allocated half-light radii $0.2 \leq r_{1/2} \leq 1.6$~kpc, according to the $z=5$ size--luminosity relationship of \citet{Bouwens2004} with a small scatter ($\sigma=0.2$~kpc) around the relation.
This was implemented in the simulations by broadening the PSF with an appropriate smoothing kernel.
As discussed earlier, unknown sizes in this range, when convolved with the PSF and measured in our adopted apertures, imply errors on $\beta$ of $<0.2$.
We assume a wavelength-independent morphology over the fairly narrow rest-frame wavelength range of interest ($\lambda\sim$1300-3000~\AA).
 
\subsection{Simulation pipeline}
The model galaxies were inserted into the images, avoiding existing sources by use of the segmentation map. 
In the HUDF, 50 copies of the field were used to avoid excess crowding.
The simulated galaxies were observed in the same way as in the real data.
While no `stars' were injected into the simulations, we performed the same star-rejection routine as for the data such that its effect on the selection efficiency could be determined.

\subsection{Selection efficiency}
\begin{figure}
\includegraphics[width=3.4in]{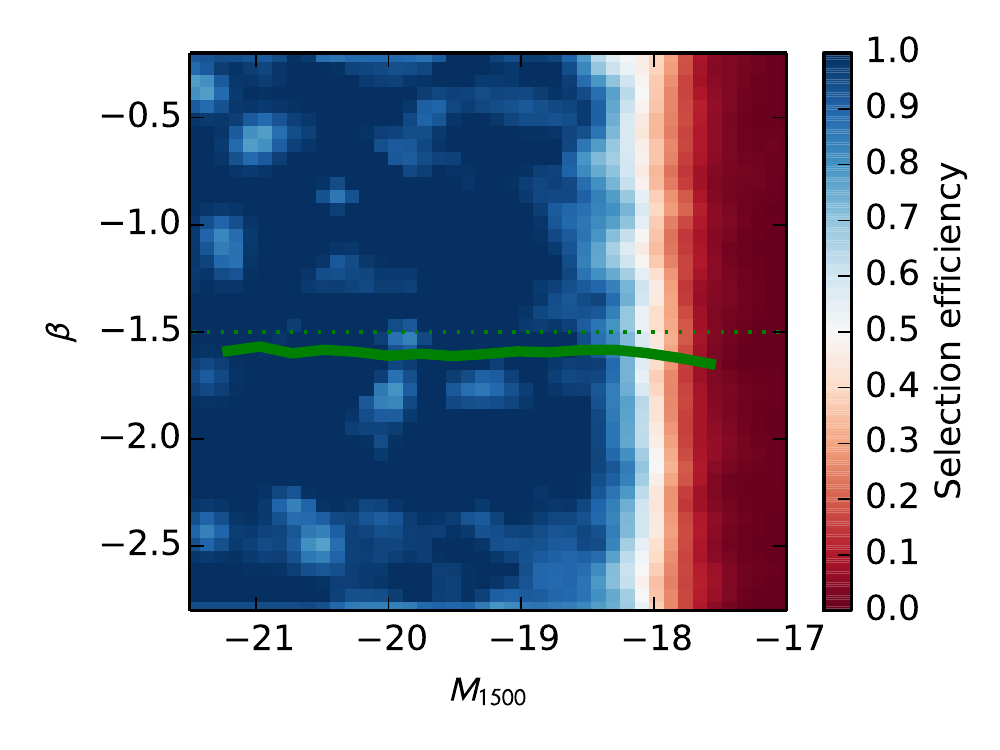}
\caption{The selection efficiency (recovered fraction of total input galaxies) is shown in bins of colour--magnitude space for the HUDF simulation.
The completeness limit lies at a fairly uniform magnitude across the range of $\beta$s, meaning no colour bias exists in any luminosity bin.
As an example we show the worst-case scenario, picking a $\beta=-1.5$ sample from the simulations (dotted green line) which, after observation, is measured as the biased solid green line.
This line is typical for colours $\beta>-2$; the bias line is even flatter for $\beta<-2$.
This small offset is (the worst) example of the systematic offsets from the simulations, $\Delta\beta\lesssim0.1$, due to the way galaxy templates deviate from a perfect power law.
The bias line is truncated at the faint limit of our sample.}
\label{fig:efficiency}
\end{figure}

Fig. \ref{fig:efficiency} shows an example of the selection efficiency of galaxies in colour--magnitude space from our HUDF simulation.
Reassuringly, considering our conservative approach, there is little preferential selection of galaxies at any colour: the completeness limits are $\beta$ independent.
Our simulations do however show evidence of some systematic offsets, although even in the worst-case scenario these are at the $\delta\beta\lesssim0.1$ level (see Fig. \ref{fig:efficiency}).
The remaining bias is due to a combination of aperture/size effects and variations in $\beta$ which depend on the filter-set or measurement method used \citep[see fig. 2 of][]{Rogers2013}.

\subsection{Comparison to data}
By multiplying the selection efficiency of the appropriate simulation by the area of each field and integrating down the adopted luminosity function model, a predicted luminosity distribution for our sample can be computed.
For this, we adopt the $z=5$ luminosity function determination of \citet{McLure2009}.
Since the luminosity function used for the simulation inputs is essentially `divided out' of this calculation (via the selection efficiency fraction), we are free to adopt this alternative luminosity function for this comparison.
Over the narrow redshift range of our selection window, the fixed-redshift function of \citet{McLure2009} provides a better representation of the luminosity distribution of galaxies than does the simple evolving parametrization used earlier: the deviation is up to a factor of 1.5 at the bright end of our sample at $z=5$.
The resultant predicted luminosity distribution is shown as the grey line in Fig. \ref{fig:cmr}, and agrees well with the combined luminosity distribution of the actual sample.
The simulations also predict uncertainties on individual galaxy $\beta$ measurements which are very similar to those found for the sample, as shown in Table \ref{tab:sample}.

\section{Intrinsic variation}
\label{sec:intrinsicvariation}

\begin{figure*}
\includegraphics[width=6.5in]{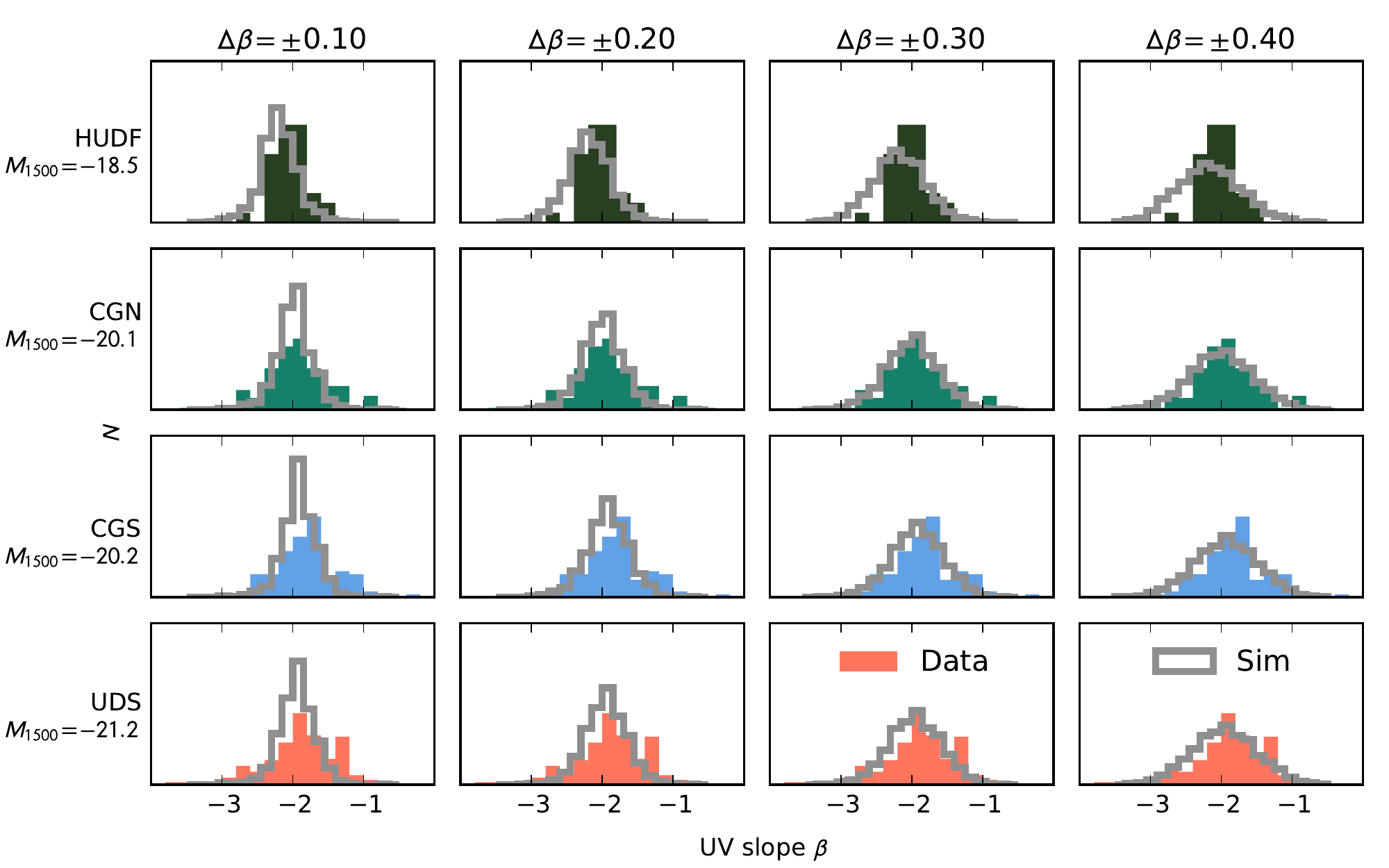}
\caption{A comparison of the intrinsic scatter of faint galaxies (top panels, from the HUDF sample) through to brighter galaxies (lower panels from CGN, CGS, and the UDS).
By comparing the distribution of colours in the data (coloured histograms) to those of the simulations (grey lines), which are designed to replicate the photometric scatter of the data, we can disentangle the intrinsic distribution of colours.
Left to right, simulations of increasing intrinsic colour variation are shown superimposed onto the fixed data.
For simplicity, we show only four possibilities here: $\Delta\beta\in\{0.1, 0.2, 0.3, 0.4\}$.
The fields containing brighter galaxies are better represented by simulations with larger intrinsic colour scatter.}
\label{fig:intrinsichists}
\end{figure*}

In this section we derive the \emph{intrinsic} variation in colour across the sample, using the image simulations to decouple the observed $\beta$ distributions into intrinsic variation and photometric scatter.
We have approached this problem from three angles.
First and most simply, we assumed that the observed $\beta$ distribution is a convolution of two Gaussians: one representing the intrinsic colour distribution of the galaxy population, and one due to photometric scatter in some fiducial scenario.
Second, by drawing realisations from the simulations according to various intrinsic distributions, we compared the data and simulations using an equal-variance test and maximized the probability that the data and realisation are from the same population.
Third, the comparisons of the $\beta$ distributions of the data and simulations were made by a full maximum-likelihood test.
Each method relies on a comparison of the observed distribution to some subset of our simulations.
A visual comparison of this is given in Fig. \ref{fig:intrinsichists}, where the observed $\beta$ distribution in each field is shown alongside simulated distributions based on various intrinsic colour scatters. 
The intrinsic distribution is in all cases assumed to be Gaussian, and is parametrized by $\Delta\beta$, its standard deviation.
We have tested a log-normal distribution and, like \citet{Castellano2012}, find no convincing preference for it.

\subsection{Measuring $\bmath{\Delta\beta}$: Gaussian assumption}
The simplest estimate of the intrinsic distribution of colours is to assume that the observed distribution is a combination of two Gaussian distributions: one reflecting intrinsic variation and one due to photometric scatter.
Under this assumption, the width of the intrinsic colour distribution is
\begin{equation}
\Delta\beta\approx\sqrt{\sigma_{\rm obs}(\beta)^2 - \sigma_{\rm photo}(\beta)^2},
\label{eqn:sqrtscatter}
\end{equation}
where $\sigma_{\rm photo}(\beta)$ can be measured by looking at the simulated $\beta$ distribution of a set of galaxies that were input with $\beta_{\rm input}\approx{\rm median}(\beta_{\rm data})$.
By relying on the varying average luminosity of galaxies in each field, we can make these comparisons along the luminosity baseline without the difficulties of combining the data and simulations of the different survey fields.
To better populate the luminosity space, each field and simulation were split into three luminosity bins of equal occupancy.
The results of this measurement are shown as $\times$ marks in the first panel of Fig. \ref{fig:scattermagnituderelation}.
The colours of the brighter galaxies found in the UDS and CANDELS fields populate a much broader distribution than photometric scatter alone would predict: hence, they stem from a more intrinsically varied population.

\subsection{Measuring $\bmath{\Delta\beta}$: equal-variance test}
\begin{figure*} 
\includegraphics[width=7.0in]{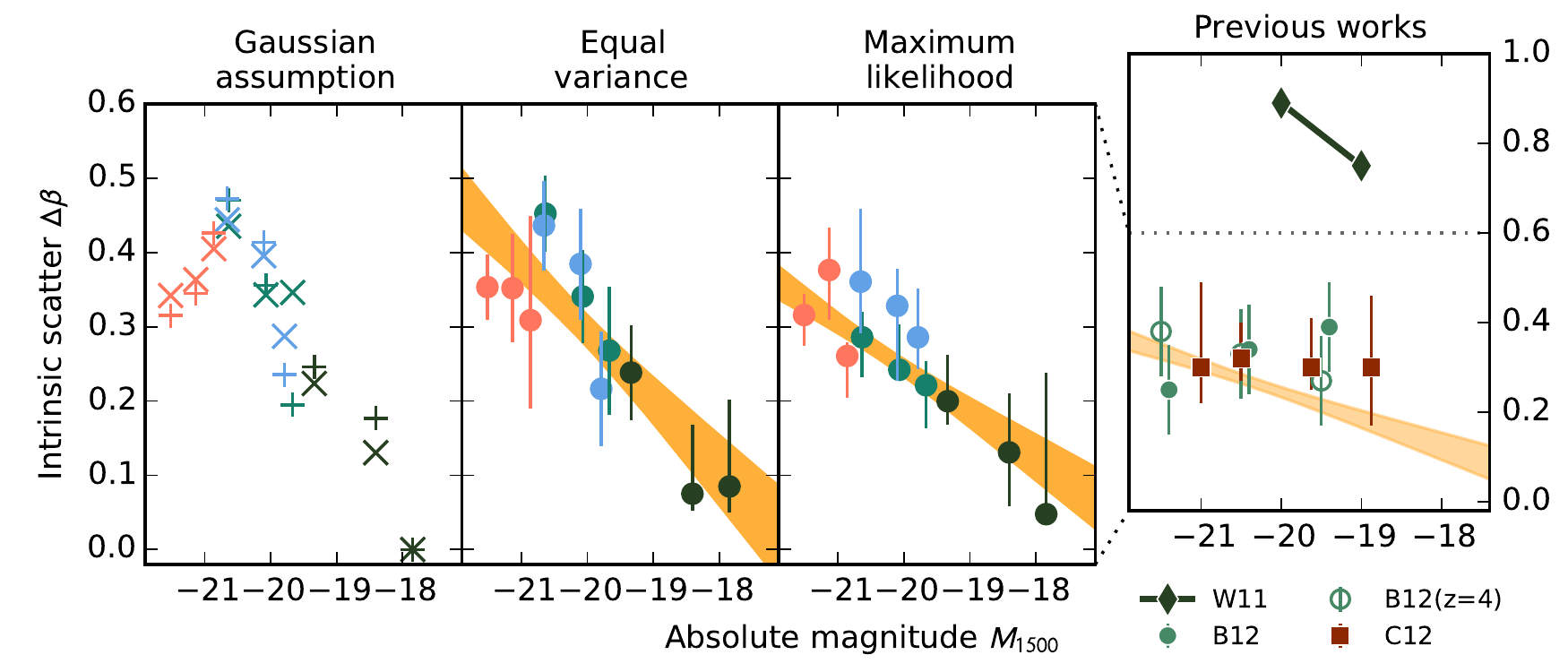}
\caption{The width of the intrinsic colour distribution of galaxies at various luminosities.
The first three panels relate to each of our three test methods, as denoted by the panel title and described in Section \ref{sec:intrinsicvariation}.
In the first panel, $\times$ marks show the results of our Gaussian assumption test, while $+$ marks denote our simulation-free check results.
In all cases, each field contributes three bins of equal occupancy.
The points are coloured by field, as in Fig. \ref{fig:intrinsichists}: salmon=UDS, blue=CGS, green=CGN, dark=HUDF.
The error bars each enclose 68\% of the total $p(\Delta\beta)$ where applicable.
In each case, the yellow regions show the $1\sigma$ error limits of a linear fit to the measurements, weighted by the errors on $\Delta\beta$.
In all three tests, brighter galaxies are drawn from a more varied intrinsic population than fainter galaxies.
The final panel (with an expanded vertical range) shows measurements drawn from the literature over narrower dynamic ranges, from \citet[B12]{Bouwens2012}, \citet[W11]{Wilkins2011}, and \citet[C12]{Castellano2012}.
The maximum likelihood relation is shown in pale yellow in the final panel for comparison.
}
\label{fig:scattermagnituderelation}
\end{figure*}

By testing against only a single simulation, the first test could not measure the uncertainty on $\Delta\beta$.
So for the equal-variance test a grid of simulated $\beta$ distributions, each according to a different $0<\Delta\beta<1$, was created for each field.
In each case, the simulation was centred on the median $\beta$ of the data.
Centring the distribution in this way, rather than around the linear CMR, avoids making any \emph{a priori} assumption about the shape of the CMR.
Following \citet{Dunlop2013}, we used a non-parametric test to assess the probability that each bin's simulated distribution and data arose from the same population.
Since for this measurement the mean intrinsic value of $\beta$ is not of interest, the \citet{Brown1974} test for equal variances, rather than a K-S test, was adopted.
In this manner, a probability density function $p(\Delta\beta)$ was created for each of the luminosity bins.
By finding the maxima of $p(\Delta\beta)$, and the regions of $\Delta\beta$ enclosing 68\% of $p$, robust measurements for $\langle\Delta\beta\rangle$ and its uncertainty were found.
Since the actual subset of simulated galaxies returned is random (in order to approximately populate a Gaussian in intrinsic $\beta$), we averaged the best value and uncertainties over multiple realisations at each $\Delta\beta$.
The variation in $\langle\Delta\beta\rangle$ between realisations was always much smaller the the error derived from $p(\Delta\beta)$.
The results are shown in the second panel of Fig. \ref{fig:scattermagnituderelation}, and are in excellent agreement with the first test.

\subsection{Measuring $\bmath{\Delta\beta}$: maximum-likelihood test}
For this final test, simulations were created for a grid of $\Delta\beta$ as before.
However in this case, each luminosity bin's data and simulation were binned in $\beta$ to form histograms with $N_B$ bins spanning $-4<\beta<0$.
We compared the simulated and actual histograms of $\beta$ and maximized 
\begin{equation}
\mathcal{L}=\prod_i^{N_B}\frac{s_i(\Delta\beta, \mu_\beta)^{d_i}\exp[-s_i(\Delta\beta, \mu_\beta)]}{d_i!},
\end{equation}
where $s_i$ and $d_i$ denote the simulated and actual (data) number of galaxies in the $i$th bin, and $s_i$ depends on $\Delta\beta$ and the centre of the Gaussian distribution from which the simulation is drawn, $\mu_\beta$.
In each case the normalisation of the histograms was fixed to ensure $\sum_{i}d_i / \sum_{i}s_i=1$, and the central colour $\mu_\beta$ of the simulation was allowed to float and was marginalised over.
The simulation realisation was also marginalised out.
The maximum-likelihood (ML) results are shown in the third panel of Fig. \ref{fig:scattermagnituderelation}, and are in good agreement with both of the previous tests.

\subsection{Measuring $\bmath{\Delta\beta}$: simulation-free check}
The three methods above all rely on image simulations.
To avoid complete reliance on these simulations, we repeated the Gaussian assumption test in a simulation-free way.
For each luminosity bin, we created $z$- to $H$-band observer-frame photometry for test galaxies, all with $\beta=-2$ but using the $M_{\rm UV}$ distribution of the data.
Using the median image depths given in Table \ref{tab:fields}, we perturbed the photometry with appropriately scaled Gaussian random numbers.
The UV slope was then measured using a simple power-law SED fit to the generated photometry, using the same set of filters as for the data.
This process yields a measurement of the expected photometric scatter on $\beta$ which depends only on the image depths (and an assumed fiducial value of $\beta=-2$).
Using equation \ref{eqn:sqrtscatter} again to compare to the observed data gives excellent agreement with our earlier Gaussian assumption test.
These simulation-free results are shown as the $+$ marks in the first panel of Fig. \ref{fig:scattermagnituderelation}.

\subsection{Discovery of significant colour scatter}
In all but the faintest bins considered, each test shows significant evidence of intrinsic colour scatter, $\Delta\beta>0$.
To quantify the significance of this, we compared a null hypothesis, where all luminosity bins contain intrinsic colour variation in line with the faintest bin ($\Delta\beta=0.1$), to an alternative hypothesis, where $\Delta\beta$ grows with luminosity ($\Delta\beta=s\times M_{1500}+c$), motivated by the trend of the ML results.
Using a maximum-likelihood estimated linear relation, the likelihood ratio test statistic
\begin{equation}
D=-2\ln\left( \frac{\mathcal{L}(\Delta\beta=0.1)}{\mathcal{L}_{\rm linear}} \right) \approx 150.
\end{equation}
Since the linear fit has two extra degrees of freedom ($s,c$), this equates to a significance of $\approx12\sigma$.
We have therefore discovered very significant intrinsic colour variation in our sample of $z\approx5$ galaxies.

\subsection{A colour-scatter--magnitude relation?}
In all of the tests considered, there is clear evidence for the intrinsic variation in colour being not only non-zero, but increasing at brighter luminosities (just as the average $\langle\beta\rangle$ is redder for brighter galaxy populations).
To measure the significance of the trend quantitatively, linear fits, weighted by the uncertainties, were made for both the equal-variance test and the ML test results.
These resultant `colour-scatter--magnitude relations' are shown as the yellow regions in Fig. \ref{fig:scattermagnituderelation}.
Formally, the slopes of the linear fits differ from 0 by $\approx4.8\sigma$ and $5.1\sigma$ for the equal-variance and ML tests respectively.
As discussed in Section \ref{sec:interlopers} regarding the CMR, removing potential lower-redshift interlopers has negligible effect on these results: in doing so the slopes then differ from 0 by $4.8\sigma$ is both tests.
As a further test of the significance, we computed a likelihood ratio of two hypotheses:
that the colour variation is a constant at all luminosities $\Delta\beta(M)=0\times M_{1500} + c=c$, or that it grows with increasing luminosity $\Delta\beta(M)=s\times M_{1500}+c$.
Taking the maximum likelihood under each hypothesis, the likelihood ratio test statistic
\begin{equation}
D=-2\ln\left( \frac{\mathcal{L}_{\rm flat}}{\mathcal{L}_{\rm relation}} \right) = 6.7,
\end{equation}
which, with the slope $s$ being one degree of freedom, makes the growing relation more likely than the flat relation at the $2.6\sigma$ level.
This is a more conservative approach and result than the linear-fit derived significance, but the two approaches are in reasonable agreement.
In all cases, the existence of a scatter--magnitude relation is confirmed at better than $2.5\sigma$ significance.

\subsection{Effect of varying signal-to-noise thresholds}
The four fields from which our sample has been assembled were treated with different SNR thresholds when selecting galaxies.
An obvious concern is that this choice may affect our results.
However, we can be confident this is not the case for a number of reasons.
Firstly the photometric uncertainty on measuring $\beta$ does not scale directly with the $z-$band detection threshold, since $\delta\beta$ depends on the noise in all the bands from which $\beta$ is measured.
Second, the simulated galaxies inhabit the same noise as the data, since we injected sources into the real images rather than merely perturbing a photometry catalogue according to some noise parametrization.
As such they are subject to the same photometric scattering and SNR cuts as the data, so the comparison of data to simulations is `fair'.
Third, the trend for brighter bins to show more intrinsic scatter is visible within each of the fields, as well as between fields.
Finally, if the simulations were systematically underestimating $\delta\beta$ then lower-SNR selected fields (i.e. the HUDF) would require more, not less, intrinsic scatter to match the observed distribution than the higher-SNR selected fields.

\subsection{Simulation stellar population and dust law choices}
While other stellar population models could have been chosen for the simulated galaxies, each model having a different mapping between $E(B-V)$, $t$ and $\beta$, these would have little effect on the actual photometry for a given $\beta$.
This is because the rest-frame UV continuum of young star-forming galaxies is in most cases well represented by a power law \citep{Leitherer1999}.
The primary exception to this would be the existence of a 2175~\AA\ dust bump in the UV attenuation curve \citep[e.g.][]{Fitzpatrick1986}, breaking the otherwise power-law like SED.
We have tested the effect such a feature would have on our $\beta$ measurements by creating photometry for $z\approx5$ star-burst galaxies using the modified Calzetti dust law of \citet{Massarotti2001}, which includes a 2175~\AA\ dust bump of amplitude $\alpha=0.25$ (calibrated by $z\approx2$ star-burst galaxies).
At $4.5<z<5.5$, the bump (if it exists at these early cosmic times) moves through the $J$-band.
In the extreme scenario of $\beta$ being constrained only by a $z-J$ colour (for instance if a galaxy in the UDS sample happened to lie in a particularly shallow region of the $H$-band image), galaxies having dust attenuation of $A_V=0.5$ would be measured with a redshift-dependent colour bias of $\delta\beta\leq0.2$ over the $4.5<z<5.5$ interval.
Assuming this extreme scenario, the adoption of a Massarotti dust law in place of a Calzetti law in our simulations would have lead to $\Delta\beta\lesssim0.1$ additional colour-scatter in the simulated galaxies' photometry.
In reality, the bias will almost always be less severe: photometry through \emph{HST}'s wider $J_{125}$ filter is less affected by the bump than UKIRT's $J$ band, and the biased $z-J$ colour is always tempered by the use of other bands ($Y$ and $H$) in the power-law fitting of $\beta$.
Overall, it is expected that the $\Delta\beta$ results will be affected by dust-law choice by $\ll0.1$.

\subsection{Comparison to previous works}
Estimates of $\Delta\beta$ at $z=4$ and $z=5$ have been published by \citet{Castellano2012}, \citet{Bouwens2012} and \citet{Wilkins2011}, but each covers only a narrow range of luminosity.
The fourth panel of Fig. \ref{fig:scattermagnituderelation} includes these previous estimates alongside our results.
Individually the literature results show little evidence of luminosity trends; in particular the trend reverses between the $z=4$ and $z=5$ samples of \citet{Bouwens2012}, suggesting that the luminosity dependence of $\Delta\beta$ is poorly constrained in both cases 
(in line with their suggested uncertainties of $\pm0.1$).
Still, within their errors, the results of \citet{Bouwens2012} and \citet{Castellano2012} are statistically consistent with our relation.

Unlike the power-law approach of \citet{Castellano2012}, \citet{Bouwens2012} and this work, \citet{Wilkins2011} chose to use a single, short-baseline $z_{850}-Y_{105}$ colour to estimate the UV continuum slope of $z\approx5$ galaxies.
In comparing their results (which were presented in observed colour terms) to ours, a transformation $\beta=6.21(z_{850}-Y_{105})-2$ is required, where the large multiplier on the observed photometry is due to the proximity of the $z$ and $Y$ filters to one another in wavelength.
However, doing so assumes, perhaps unreasonably, that the intrinsic variation in the rest-frame [1500~\AA]$-$[1730~\AA] colour traces the variation in the rest-frame [1500~\AA]$-$[2540~\AA] colour (approximately the original definition of $\beta$ by \citet{Calzetti1994} and spanned by $z_{850}-H_{160}$ at $z=5$).
Still, to draw a comparison to the results of \citet{Wilkins2011} we differenced their quoted observed and simulated $z-Y$ colours using equation \ref{eqn:sqrtscatter} and multiplied the inferred intrinsic colour scatter by 6.21, yielding surprisingly large estimates of $\Delta\beta\approx0.8$ at $M_{1500}\approx-19.5$.
The discrepancy between this result and the other literature values (and ours) can be ascribed to the much larger uncertainty in $\beta$ provided by the $z-Y$ colour, and the shorter wavelength range it probes.
As quantified by \citet{Bouwens2012} in their appendix B.3, the use of a UV-spanning $z-H$ colour, rather than a $z,Y,J,H$ power-law fit, increases the uncertainty in $\beta$ by a factor 1.5.
The impact is even more dramatic with the narrow $z-Y$ colour: measuring $\beta$ in this way for our UDF sample yields a standard deviation $\sigma(\beta_{z-Y})=1.32$, as opposed to this work’s $\sigma(\beta_{\rm power-law})=0.26$.
Overall this is unsurprising since, aside from exploiting three additional filters, $\beta_{\rm power-law}$ is constrained by a wavelength baseline $4.5\times$ longer than that of the $z-Y$ colour.
Of course, the simulations of \citet{Wilkins2011} were treated with the same colour measurement as their data, meaning $\Delta\beta$ should still be inferable, albeit with a large uncertainty.

In summary, our relation, which constitutes the first significant measurement of the luminosity dependence of $\Delta\beta$, is not in contention with results from the previous studies which shared our approach to measuring $\beta$.
Our measured level of colour-scatter (over a wide rest-frame UV baseline) is notably lower than the intrinsic scatter which \citet{Wilkins2011} find using a single $z-Y$ colour, just above the Lyman break.
This is due in part to increased uncertainties in their colour, and perhaps also to local variations in the SED at those wavelengths compared to the broader rest-frame UV.

\subsection{Asymmetric colour scatter}
\begin{figure} 
\includegraphics[width=3.2in]{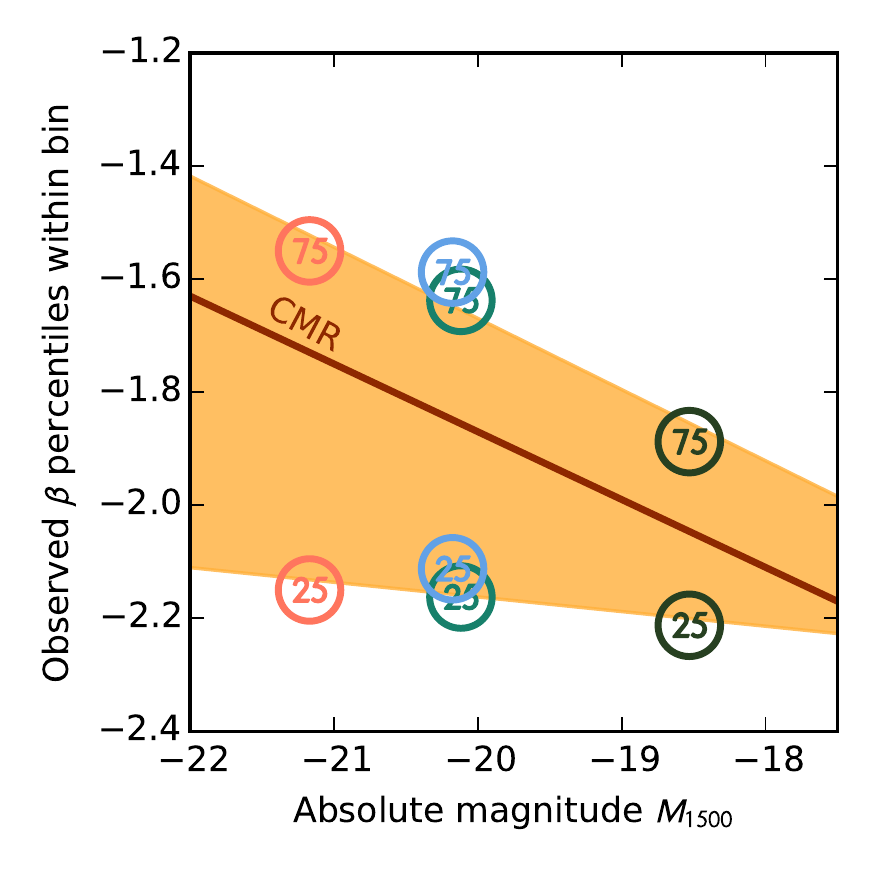}
\caption{The 25th and 75th percentiles of $\beta$ in each field are shown by coloured circles (marked 25 and 75, respectively).
The edges of the shaded region are linear fits to the points, weighted by errors determined via bootstrapping.
The region thus denotes a simple measure of how the scatter in $\beta$ increases to bright luminosities as in Fig. \ref{fig:scattermagnituderelation}.
Here however, it is clear that while the bluest galaxies are similarly blue at all luminosities, the redder average colours and greater colour scatter at bright $M_{1500}$ are driven by establishing a redder sub-population in the brighter bins.
The red line is the average colour-magnitude relation from Fig. \ref{fig:cmr}.
Otherwise, the field colouring matches that of Fig. \ref{fig:intrinsichists}: salmon=UDS, blue=CGS, green=CGN, dark=HUDF.
}
\label{fig:redorblue}
\end{figure}
Is the colour--magnitude relation itself merely a product of the scatter--magnitude relation?
For this to be true, the scatter must expand in such a way that brighter bins include more red galaxies than fainter bins without many more blue galaxies.
Fig. \ref{fig:redorblue} offers some evidence that this may be the case.
Between the four fields, the 25th percentiles of the observed $\beta$ distribution grow no bluer ($\beta\approx-2.1$) from the faintest HUDF galaxies to the brightest UDS galaxies.
Meanwhile the 75th percentiles redden from $\beta=-1.9\rightarrow-1.5$.
The bluest galaxies are not any bluer in the brightest bins, but the reddest galaxies are much redder.
A distribution which grows to the red would be in line with the `blue-ridge--red-tail' distribution of blue-sequence field galaxies in the $z\lesssim3$ study of \citet{Labbe2007}.
This is perhaps an unsurprising discovery: galaxy formation models predict the dust-free colours of high-redshift galaxies to follow only a weak luminosity dependence, and be $\beta\gtrsim-2.5$ at $z\approx5$ \citep[e.g.][]{Wilkins2013}. 
But with no strict upper limit to their dust reddening, the colour--magnitude relation may simply follow the typical dust reddening at each UV luminosity.
In this scenario, the colour--magnitude relation is driven by the increasing fraction, with increasing luminosity, of galaxies which are reddened.
Alternatively, the asymmetry may be driven by the ability for more luminous galaxies to harbour somewhat older stellar populations, built up over longer periods of time, than fainter galaxies.
In this scenario, UV luminosity traces stellar mass and more massive galaxies are observed at various stages of their star-formation duty cycle: the red wing represents those galaxies with the longest period of quiescence since their last star burst.
Less massive galaxies, at the faint end of the CMR, are all currently star forming and do not harbour a significant older stellar population.


\section{Conclusions}
The rest-frame UV colours of high-redshift galaxies provide a probe of the metallicity and dust conditions within which their stellar populations are growing.
While recent attention has been focused at $z\geq7$, moving just 400~Myr later to $z=5$ has allowed us take a more conservative approach to the detection, selection, and colour analysis of a sample of $z\approx5$ Lyman-break galaxies.
Detected mostly at SNR~$>8\sigma$, and with four or five imaging bands spanning the rest-frame UV, we have been able to robustly determine the UV continuum slope $\beta$ of each of 584 $z\approx5$ LBGs (typically to better than $\delta\beta\lesssim0.4$).
Crucially these galaxies span a factor of 100 in luminosity, allowing us to constrain the colour--magnitude relation.
Comparing our samples to closely representative image simulations of mock galaxies, we have also disentangled the intrinsic variation in colour at each magnitude from the photometric scatter.
Our findings can be summarised as follows.
\begin{enumerate}
\item{A linear colour--magnitude relation, whereby brighter galaxies are redder than fainter galaxies by ${\rm d}\beta/{\rm d}M_{\rm UV}=-0.12\pm0.02$ provides a good fit to our data.}
\item{The data show no convincing evidence either for or against a piecewise-linear relation, whereby galaxies cease to get bluer with decreasing luminosity below some point, as had been suggested elsewhere.}
\item{For the first time over a wide range of luminosities at high redshift, we have discovered significant ($12\sigma$) evidence for \emph{intrinsic colour variation} within the LBG population.}
\item{This intrinsic colour variation is significantly larger ($2.5\sigma$) in high-luminosity bins than low-luminosity bins, after accounting for photometric scatter using our detailed simulations. This result was confirmed by multiple statistical tests, as well as by a final check which was not reliant on our simulations in any way.}
\item{The luminosity-dependent colour scatter and average colour--magnitude relation appear to be due to the evolution of bright, red galaxies.
This appearance of bright, red galaxies coincides with a seemingly luminosity-independent blue floor: in each luminosity bin, the 25th percentiles of colour are always $\beta\approx-2.1$, while the 75th percentiles grow redder from $\beta=-1.9\rightarrow-1.5$ over $M_{1500}=-18.5\rightarrow-21.2$.}
\end{enumerate}

The rest-frame UV colour $\beta$ is dependent on all parameters of the stellar population, but is particularly sensitive to light-weighted age and dust attenuation.
Our measurements of lower-luminosity galaxies, $\langle\beta(M_{1500}=-18)\rangle\approx-2$, are not so blue as to require dust-free stellar populations.
However the lack of intrinsic scatter there ($\Delta\beta<0.2$) shows that, if the galaxies are dust reddened, it is by similar amounts for all galaxies in that bin.
Similarly the light-weighted ages of galaxies in the faint population must be fairly similar to one another, and $<100$~Myr.
This can be interpreted as all of those galaxies undergoing intense present or recent star-formation.
Comparatively, the average brighter galaxy, with colour $\langle\beta(M_{1500}=-21)\rangle\approx-1.7$, must have built an older stellar population, or have higher metallicity or dust reddening.
However at all luminosities a quarter of galaxies have colours bluer than $\beta\lesssim-2.1$, so even at $M_{1500}=-21$ the low reddening, young galaxies remain common.

These observations are consistent with at least two simple scenarios, between which our current observations cannot differentiate:
the build up of dust as the galaxy grows and brightens, with some galaxies oriented such that the UV light escapes with less than average reddening; or stochastic star-formation histories, where faint galaxies are always currently star-forming but brighter galaxies are observed during various phases of their star-formation duty cycle.

\section*{Acknowledgements}
We thank the referee for their helpful comments which greatly improved this paper.
This research made use of Astropy, a community-developed core Python package for Astronomy \citep{Astropy2013}, NumPy and SciPy \citep{Oliphant2007}, Matplotlib \citep{Hunter2007}, {IPython} \citep{Perez2007}, NASA's Astrophysics Data System Bibliographic Services and has benefited from the SpeX Prism Spectral Libraries, maintained by Adam Burgasser at \url{http://pono.ucsd.edu/~adam/browndwarfs/spexprism}.

ABR and EFCL acknowledge the support of the UK Science \& Technology Facilities Council.
RJM acknowledges the support of the European Research Council via the award of a Consolidator Grant (PI McLure).
JSD, RAAB, and PD acknowledge the support of the European Research Council via the award of an Advanced Grant to JSD, who also acknowledges the support of the Royal Society via a Wolfson Research Merit award and the contribution of the EC FP7 SPACE project ASTRODEEP (Ref. No: 312725).
This work is based in part on observations made with the NASA/ESA \emph{Hubble Space Telescope}, which is operated by the Association of Universities for Research in Astronomy, Inc., under NASA contract NAS5-26555.

\bibliographystyle{mn2e}                      
\bibliography{z5lbgcmr}       


\label{lastpage}

\end{document}